 \documentclass[smallabstract,smallcaptions]{dccpaper}

\usepackage{epsfig}
\usepackage{amsmath}
\usepackage{amssymb}
\usepackage{color}
\usepackage{url}
\usepackage{hyperref}
\usepackage[capitalise]{cleveref}

\newlength{\figurewidth}
\newlength{\smallfigurewidth}
\usepackage{graphicx} 
\usepackage{subfigure}

\usepackage{physics}
\usepackage{algorithm}
\usepackage{algpseudocode}

\begin{document}
\title
{\large
\textbf{Feature Compression for Machines with Range-Based Channel Truncation and Frame Packing}
}

\author{%
Juan Merlos$^{\ast}$$^{\dag}$, Fabien Racap\'e$^{\ast}$, Hyomin Choi$^{\ast}$, Mateen Ulhaq$^{\ast}$ and Hari Kalva$^{\dag}$\\[0.5em]
{\small\begin{minipage}{\linewidth}\begin{center}
\begin{tabular}{ccc}
$^{\ast}$InterDigital - AI Lab & \hspace*{0.5in} & $^{\dag}$Florida Atlantic University \\
\url{first.last@interdigital.com} && \url{{jmerlosjr2017,hkalva}@fau.edu} 
\end{tabular}
\end{center}\end{minipage}}
}


\maketitle
\thispagestyle{empty}

\begin{abstract}
This paper proposes a method that enhances the compression performance of the current model under development for the upcoming MPEG standard on Feature Coding for Machines (FCM). 
This standard aims at providing inter-operable compressed bitstreams of features in the context of split computing, i.e., when the inference of a large computer vision neural-network (NN)-based model is split between two devices. 
Intermediate features can consist of multiple 3D tensors that can be reduced and entropy coded to limit the required bandwidth of such transmission. 
In the envisioned design for the MPEG-FCM standard, intermediate feature tensors may be reduced using Neural layers before being converted into 2D video frames that can be coded using existing video compression standards. 
This paper introduces an additional channel truncation and packing method which enables the system to preserve the relevant channels, depending on the statistics of the features at inference time, while preserving the computer vision task performance at the receiver. 
Implemented within the MPEG-FCM test model, the proposed method yields an average reduction in rate by 10.59\% for a given accuracy on multiple computer vision tasks and datasets.
\end{abstract}

\section{Introduction} \label{sec:intro}

Larger and larger neural networks (NN) that perform complex computer vision tasks are being deployed.
These networks expect data that is often captured by computationally-limited devices.
To help cope with computational and energy requirements, the vision task computations are off-loaded to a remote device, though at the cost of additional network latency and potential unreliability. 
When the device cannot run any NN-operations due to hardware limitations, the task can fully run on a remote device, requiring the compression and transmission of the video data. This pipeline is referred to as Video Coding for Machines (VCM), and is studied in the dedicated activity at MPEG/ISO by the MPEG-VCM group. 

\begin{figure}
    \centering
    \includegraphics[width=1\linewidth]{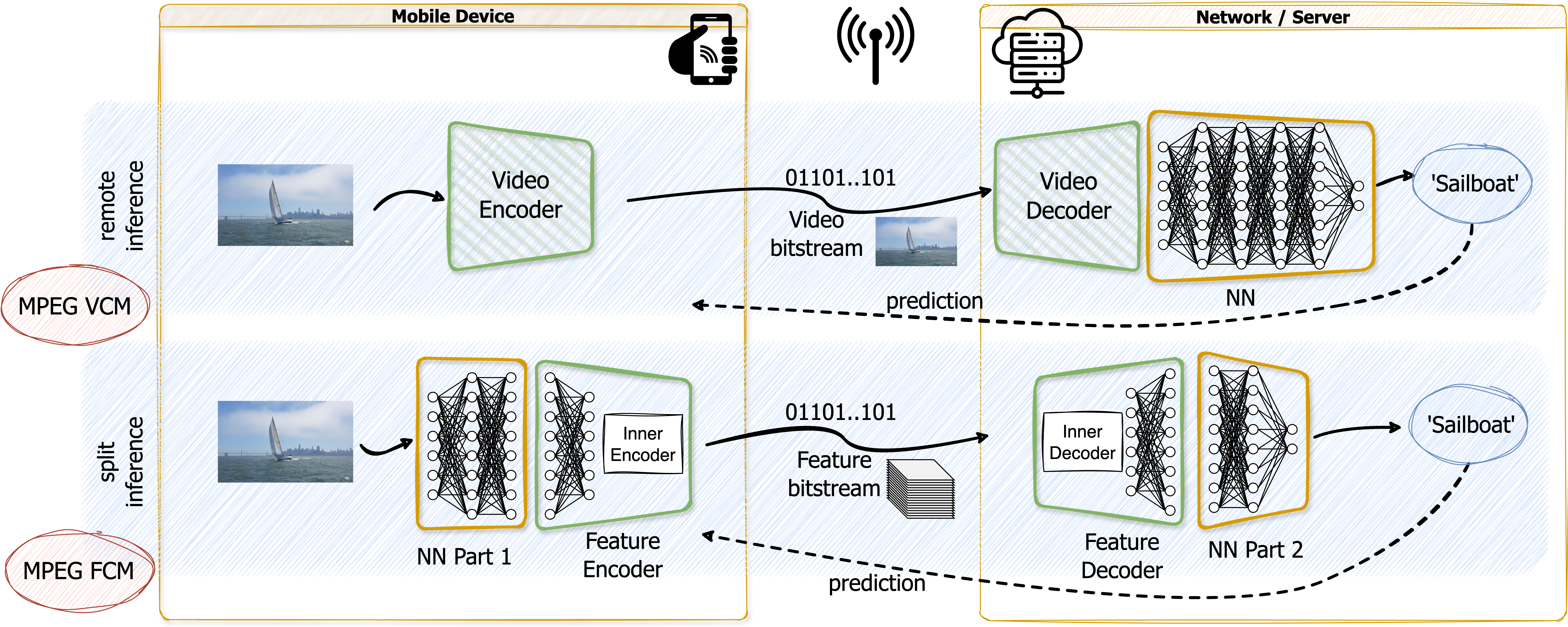}
    \caption{Types of network-assisted pipelines. Top row: remote inference with video compression. Bottom row: split-inference with feature compression}
    \label{fig:remote-split}
\end{figure}

However, the paradigm of split-computing has gained recent interest, where the inference of a neural network is shared between multiple devices~\cite{shlezinger2022collaborative}.
The intermediate data at the split point may however be too large to transmit via wireless communication systems. To overcome these limitations and enhance distributed computing, MPEG released a Call for Proposals (CfP) in 2023 for Feature Coding for Machines (FCM) technology, initially referred to as Feature Compression for Video Coding for Machines (FCVCM) \cite{fcmCFP}. 
Both VCM and FCM pipelines are represented in Figure~\ref{fig:remote-split}, where the VCM approach relies on video codecs and the receiver runs the task on decoded video content, whereas in the FCM approach, a feature codec is required to compress and transmit the intermediate data.

\begin{figure}
    \centering
    \includegraphics[width=1\linewidth]{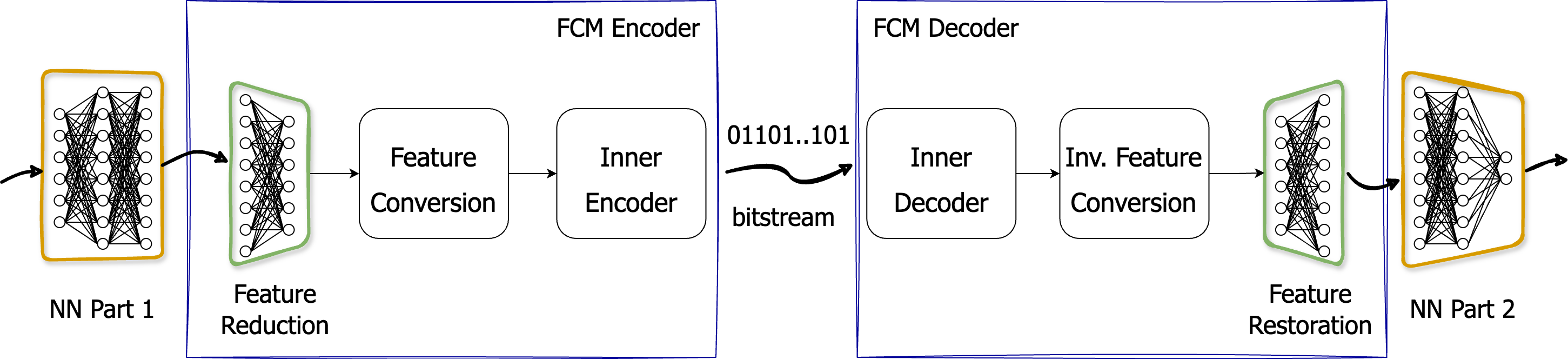}
    \caption{Feature Coding for Machines overview}
    \label{fig:codec}
\end{figure}

As depicted in Figure~\ref{fig:codec}, the current state-of-the-art FCM model employs different techniques to reduce the bitrate of the compressed bitstream of intermediate features while maintaining comparable task performances. The Feature Reduction involves condensing the original neural network split point output tensors, a process, achieved through learned reduction. Subsequently, feature conversion scales and packs the reduced feature channels into a tiled 10-bit monochrome format, which is then encoded using a traditional video codec, such as H.266/VVC \cite{vvc}. The reverse process is applied at the decoder to restore the features to their original state, enabling the receiver to run the second part of the split network and compute task predictions \cite{fcmWD}.

\begin{figure}
    \centering
    \includegraphics[width=.75\linewidth]{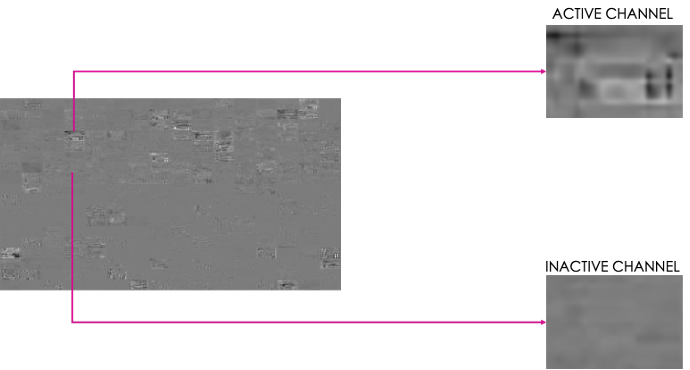}
    \caption{Example of high activation vs low activation feature channel}
    \label{fig:channel-activations}
\end{figure}

During this process, intermediate feature channels are computed that contain small, noisy activations that have minimal impact on network task performance. Despite this, the latent noise within these channels is still passed to the encoder after feature conversion. This paper addresses this issue by proposing a method to truncate noisy, low-activation feature channels post feature reduction, although similar methods could be applied to other stages of a feature encoding pipeline. Figure~\ref{fig:channel-activations} illustrates the presence of these inactive channels in the packed frame of feature channels following feature conversion.

The remainder of the paper is organized as follows.
\cref{sec:relwork} provides an overview of related work.
\cref{sec:prop} presents and discusses the proposed algorithm.
\cref{sec:results} presents experimental results and
\cref{sec:conc} concludes the paper.

\section{Related Work} \label{sec:relwork}

\begin{figure}[t]
\centering
\includegraphics[width=.95\linewidth]{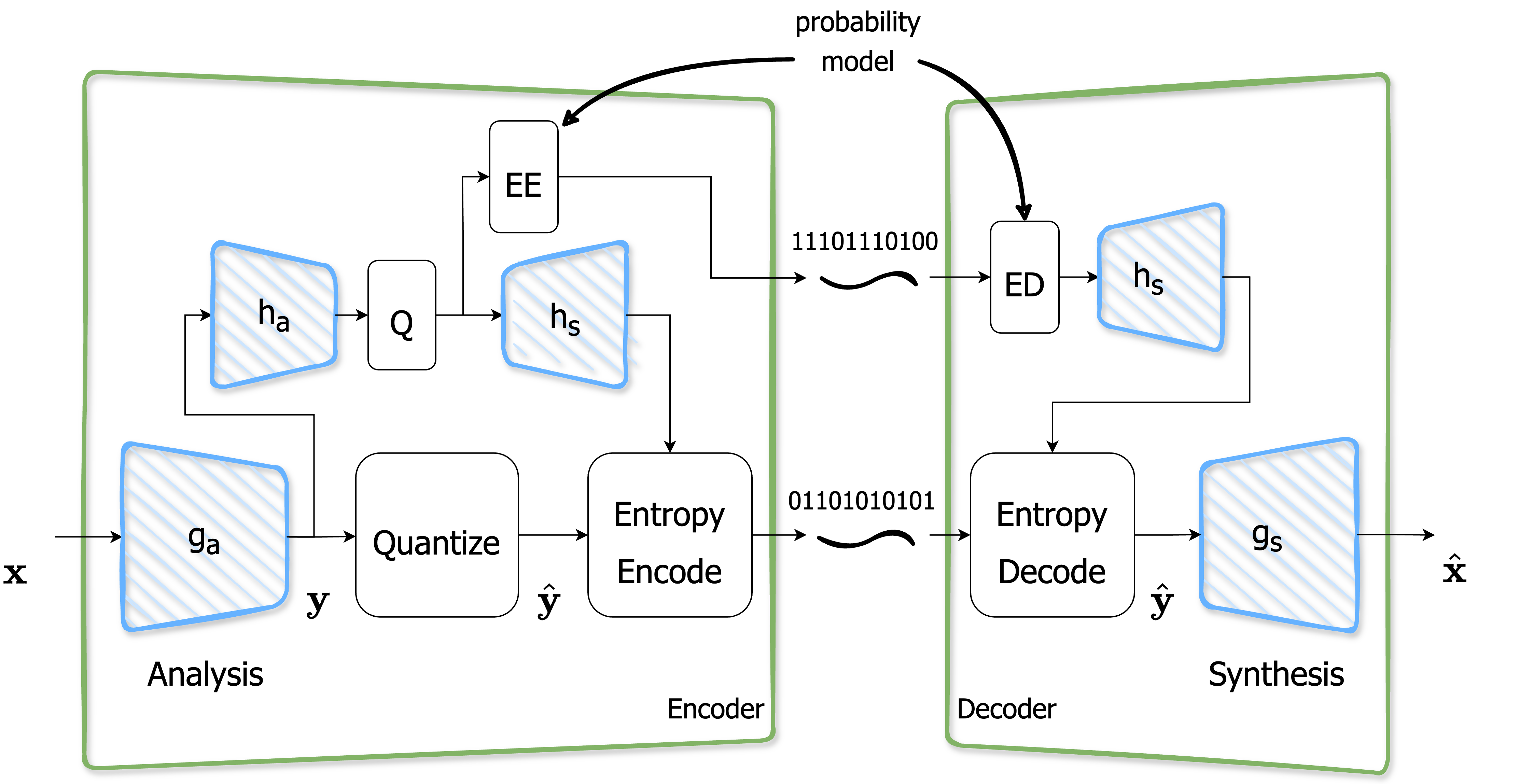}
\vspace{-0.1in}
\caption{\label{fig:scale_hyperprior}
Mean\&Scale Hyperprior-based image compression model \cite{minnen2018joint}.}
\end{figure}

One challenge faced when a typical large NN-based computer vision model is split in the middle is that the size of the intermediate data, or features, may be excessively large.
To address this, Matsubara et. al \cite{matsubara2022split} propose the insertion of an NN-based bottleneck or learned compression model to dramatically reduce the size of transmitted features. 
This bottleneck applies existing learned image compression methods, but to the intermediate features within the split model, rather than to images.
Figure~\ref{fig:scale_hyperprior} depicts the scale-hyperprior learned image compression model from \cite{minnen2018joint}. 
At the encoder side, the analysis $g_a$ transforms an input $x$ into a latent tensor $y$, which is quantized as $\hat{y}$.
The encoding probability distribution $p_{\hat{y}}$ used to encode/decode $\hat{y}$ is encoded using a hyperprior network, denoted by $h_a$ and $h_s$.
At the decoder side, $g_s$ transforms the decoded latent tensor $\hat{y}$ back to the input domain so that $\hat{x}$ is a lossy reconstruction of $x$.  
For the FCM model shown in Figure~\ref{fig:codec}, $g_a$ and $g_s$ are called the feature reduction and restoration, respectively. 
However, such a fully-learned bottleneck is relatively heavyweight, and requires significant computation, which contradicts the goal of split-computing to reduce the compute load on the sensor device. 
Specifically, when it comes to video tasks such as object tracking, existing end-to-end trained codecs are significantly outperformed by existing traditional video compression standards, both in terms of compression performance as well as computational overhead. 
Therefore, the current design of the future MPEG-FCM codec relies on the usage of existing video compression standards for faster adoption and deployment on devices already equipped with dedicated hardware for those codecs.

At the same time, techniques for neural network reduction via feature activation have predominantly focused on reducing network complexity, with less emphasis on improving compression gains of intermediate data as advances made in model pruning and complexity reduction apply within the context of channel removal. 
For instance, Ganguli et al. \cite{ganguli2024} presented a method where neurons are pruned based on their activation frequency during training, resulting in a sparse, low-rank matrix approximation. A more relevant method was proposed by Hua et al. \cite{hua2019} where they describe a pruning method that reduces computation costs in CNNs by skipping computations on less influential channel regions during inference.
However, both methods rely on training operations and data, whereas the main goal of this paper is to provide a solution that directly adapts to the input content at inference time, thus enabling a more versatile feature compression standard that can adapt to different split models and input content. 

\section{Proposed Method} \label{sec:prop}
This paper proposes a method to detect and truncate low-activation feature channels in the post-reduction output feature tensor, and signal those decisions within the bitstream such that the decoder can reconstruct the tensor in its original shape. 
This adaptive online truncation aims to reduces the bitrate of the associated bitstream while maintaining comparable task network performance.

\begin{figure}
    \centering
    \includegraphics[width=1\linewidth]{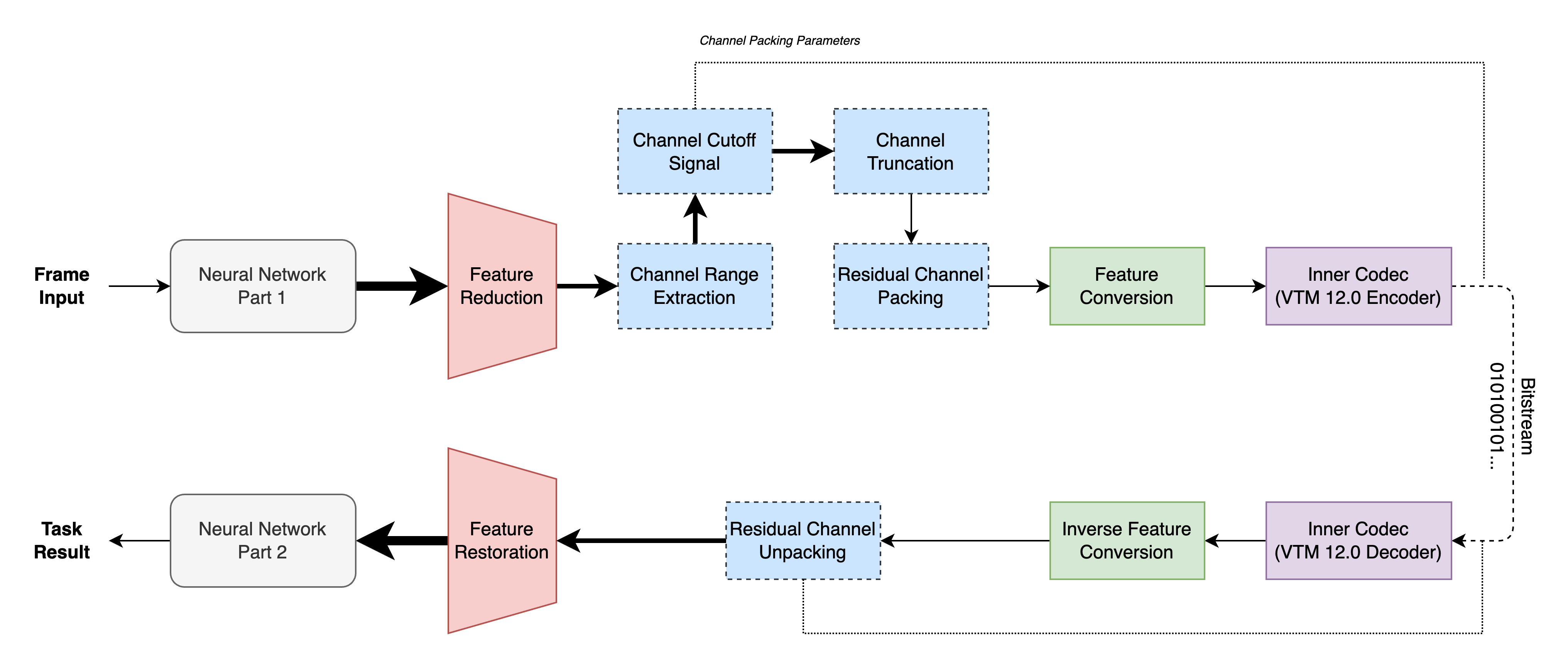}
    \caption{Proposed method overview}
    \label{fig:method_pipeline}
\end{figure}

\subsection{Channel Range Extraction and Cutoff Signal}
The range of each channel in the post-reduction tensor is calculated, representing the difference between the maximum and minimum values within that channel. Channels exhibiting a high range are considered 'active', indicating their significant potential to influence the final predictions of the task network through the activation function. In contrast, channels with lower ranges are less likely to contribute meaningfully to the network’s output. To systematically identify these less active channels, a threshold is established, channel ranges that fall below the specified cutoff threshold are marked as inactive, ensuring that only the most influential channels remain for subsequent processing.

The channel cutoff signal determines the threshold for categorizing channels as either active or inactive. This threshold is determined using the formula:
\[
\text{threshold} = \alpha \times \frac{1}{N} \sum_{i=1}^{N} \left(\max(x_i) - \min(x_i)\right), \quad \text{where } 0 < \alpha < 1
\]
where $x = \{x_1, \ldots, x_N\}$ represents the set of all channels. 
The presence of inactive channels is then defined by evaluating if their range is significant compared with the average value of the channel ranges within the tensor.
Channels that fall below this threshold are considered inactive and are selected for truncation. This automatic process adapts to the specific characteristics of the feature tensor. 
The threshold value for truncation can be adjusted at the encoder based on user's requirements: a value closer to the mean results in more channels being truncated, but also increases the risk of removing important features. 
We chose \( \alpha = \frac{2}{3} \) to strike a balance between compressing the data and preserving essential features.

\subsection{Channel Truncation}
A channel that is categorized as inactive based on the cutoff signal is removed from the set of channels available for transmission. 
This truncation process is applied only during the first frame of a given period—in our simulations, 128 frames—ensuring stability. The same set of active channels are maintained throughout this 128-frame period.
This refresh period can be signalled via FCM High Level Syntax (HLS) and can also advantageously follow existing temporal structures of the inner codec, such as the Intra Decoder Refresh period, when used.
After this period, new active channel sets are selected if needed.
The system dynamically adapts by selecting and evaluating new channel sets according to the task network's needs as determined by channel activity. 
The total number of retained channels in subsequent periods cannot exceed the initial number for the first period as it would require changing the resolution of the video encoded by the inner codec, and thus start a new bitstream.

\subsection{Residual Channel Packing}
\begin{figure}
\centering
 \subfigure[]{\includegraphics[width=.49\linewidth]{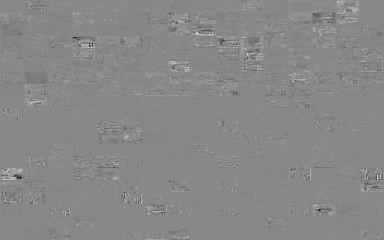}}
 \subfigure[]{\includegraphics[width=.46\linewidth]{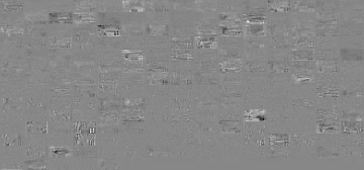}}
 \caption{Example of channel truncation and packing (b) vs. original channel packing without truncation (a)}
 \label{fig:pack_example}
\end{figure}

After truncation, the reduced set of active channels must be packed efficiently for transmission.
A new frame size is selected to accommodate the reduced channel count, optimizing the data for compression.
This packing process is crucial for bitrate reduction, as it ensures that only the essential, high-activation channels are transmitted, in addition to possibly attaining a secondary bitrate reduction by putting active channels side-by-side, particularly if similar channels happen to get placed close to each other, whereas, originally, they may have been further apart.
Figure~\ref{fig:pack_example} shows an example of the proposed packing after truncation (b) of the resulting reduced tensor obtained from the FPN backbone of the Detectron2 model~\cite{wu2019detectron2} performed on the Parkscene sequence of the SFU-HW dataset~\cite{choi2021dataset}.
Additionally, information related to the activation status of each channel is separately encoded and transmitted, enabling the decoder to accurately reconstruct the original channel structure during the unpacking process.
This information requires a maximum of one bit per post-reduction tensor channel, e.g., an overhead of 320 bits per refresh period for a tensor with 320 channels.
Although this bit-string can be entropy coded for further compression, the bitrate overhead is relatively minimal, so it is acceptable to simply transmit that information as-is alongside other HLS elements.

\subsection{Residual Channel Unpacking}
During decoding, the system unpacks the transmitted frame into its original channel structure using the activation information sent alongside the packed data. This step ensures that the original channel count is restored, allowing the feature restoration network to function as intended. The unpacking process dynamically adjusts the reconstructed channels based on the received activation signals, ensuring that the system remains efficient while preserving the integrity of the feature tensors by filling in previously truncated channels with the overall average value of maintained and transmitted feature channels, in essence creating fully flat channels to replace once noisy channels that were not transmitted.

\section{Experimental Results} \label{sec:results}

\subsection{Experimental Setup}
Our method has been implemented on top of the FCM test model (FCTM) v3.2, whose post-reduction feature tensor contains 320 channels. The results generation pipeline follows that as depicted in Figure~\ref{fig:method_pipeline}, with evaluation video datasets following that of the MPEG-FCM Common Training and Test Conditions (CTTC) \cite{fcm_cttc}, summarized in Table~\ref{tab:dataset_summary} along with evaluated task network split points.
In the context of compression for machine tasks, the compression performance trade-off is calculated as a task prediction accuracy as a function of the bitrate. 
The mean Average Precision (mAP) is used as the accuracy metric for the object detection task (SFU-HW) and Multiple Object Tracking Accuracy (MOTA) is used for tracking accuracy (TVD and HiEve datasets).
The quantization parameter QP of the inner codec corresponding to VVC's reference software VTM-12.0 is selected for each sequence of each dataset to reach 4 similar performance points as the anchor FCTMv3.2 so that bitrate variations can be compared at similar accuracy performance.
The reported Bjøntegaard Delta (BD)-rates then represent the relative variation of the bitrate at same task accuracy, over a range of accuracies, comparing the state-of-the-art system (FCTM Anchor) to the system enhanced with the proposed method. 

\begin{table}[ht]
    \centering
    \resizebox{\linewidth}{!}{
    \begin{tabular}{|l|l|l|l|l|}
        \hline
        Dataset & Task & Network & Split point & Task measure \\
        \hline
        TVD~\cite{tvd} & Object tracking & JDE-1088x608 & Darknet-53 & MOTA \\
        HiEve~\cite{hieve} & Object tracking & JDE-1088x608 & “ALT1”: layers \{75, 90, 105\}  & MOTA \\
        SFU-HW~\cite{choi2021dataset} & Object detection & FasterRCNN-X101-FPN & P-layer (P2-P5) & mAP @ 0.5-0.95 \\
        \hline
    \end{tabular}
    }
    \caption{Summary of datasets and networks.}
    \label{tab:dataset_summary}
\end{table}


\subsection{Results}

\begin{table}[ht]
\centering
\begin{tabular}{|l|c|}
\hline
\textbf{Dataset}           & \textbf{BD-Rate} \\ \hline
SFU-HW (Class A/B)            & -9.27\%                    \\ \hline
SFU-HW (Class C)              & -16.26\%                   \\ \hline
SFU-HW (Class D)              & -18.72\%                   \\ \hline
TVD (Overall)              & 0.00\%                     \\ \hline
HiEve (1080p)              & -11.11\%                   \\ \hline
HiEve (720p)               & -8.18\%                    \\ \hline
\end{tabular}
\caption{BD-Rate gains of proposed method}
\label{tab:percentage_changes}
\end{table}

\begin{figure}
\centering
 \subfigure[]{\includegraphics[width=.49\linewidth]{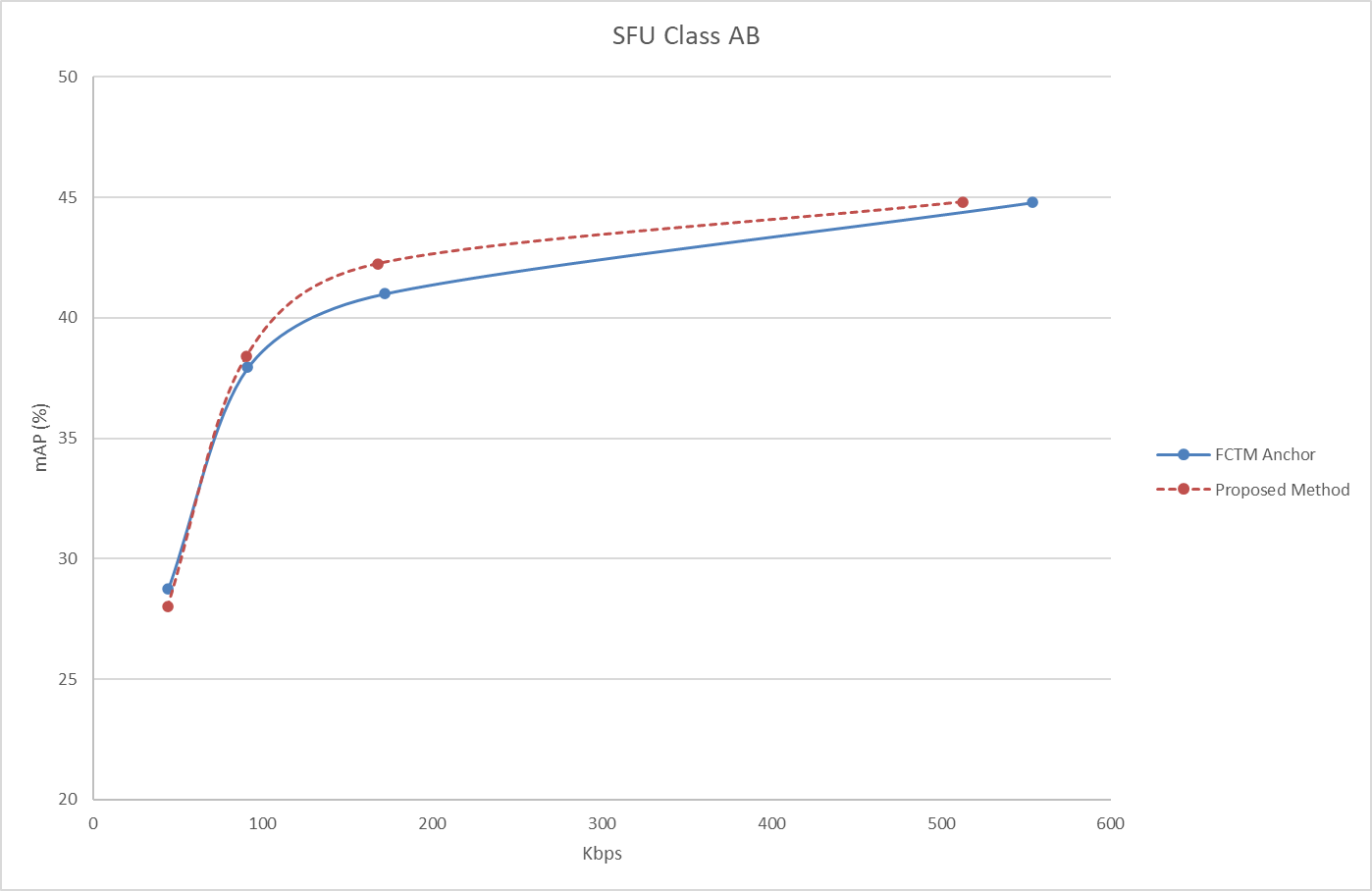}}
 \subfigure[]{\includegraphics[width=.49\linewidth]{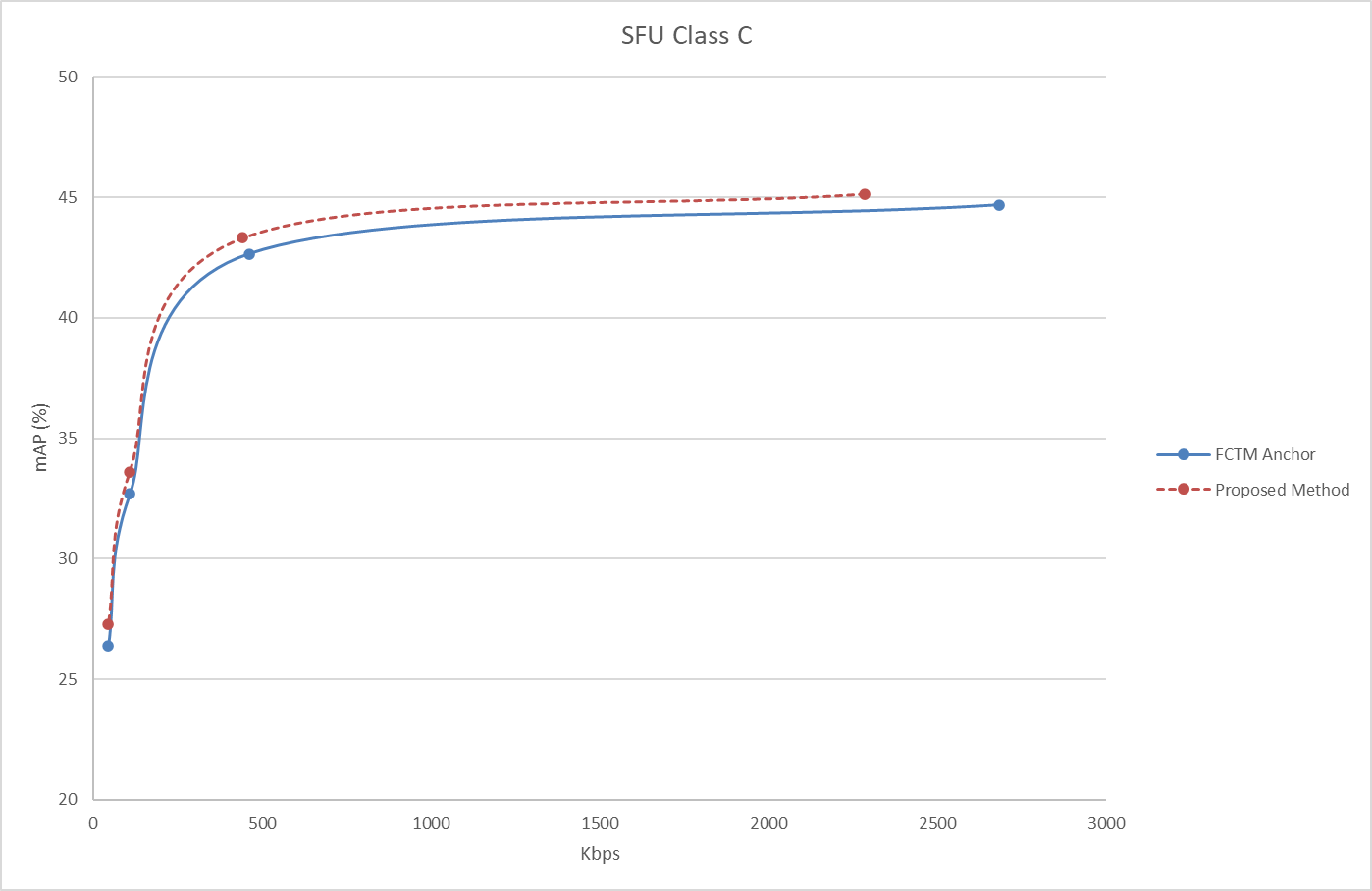}}
 \caption{SFU-HW bitrate vs. mAP curves: (a) Class A/B, (b) Class C}
 \label{fig:sfu_curve}
\end{figure}

\begin{figure}
  \centering
 \subfigure[]{\includegraphics[width=.49\linewidth]{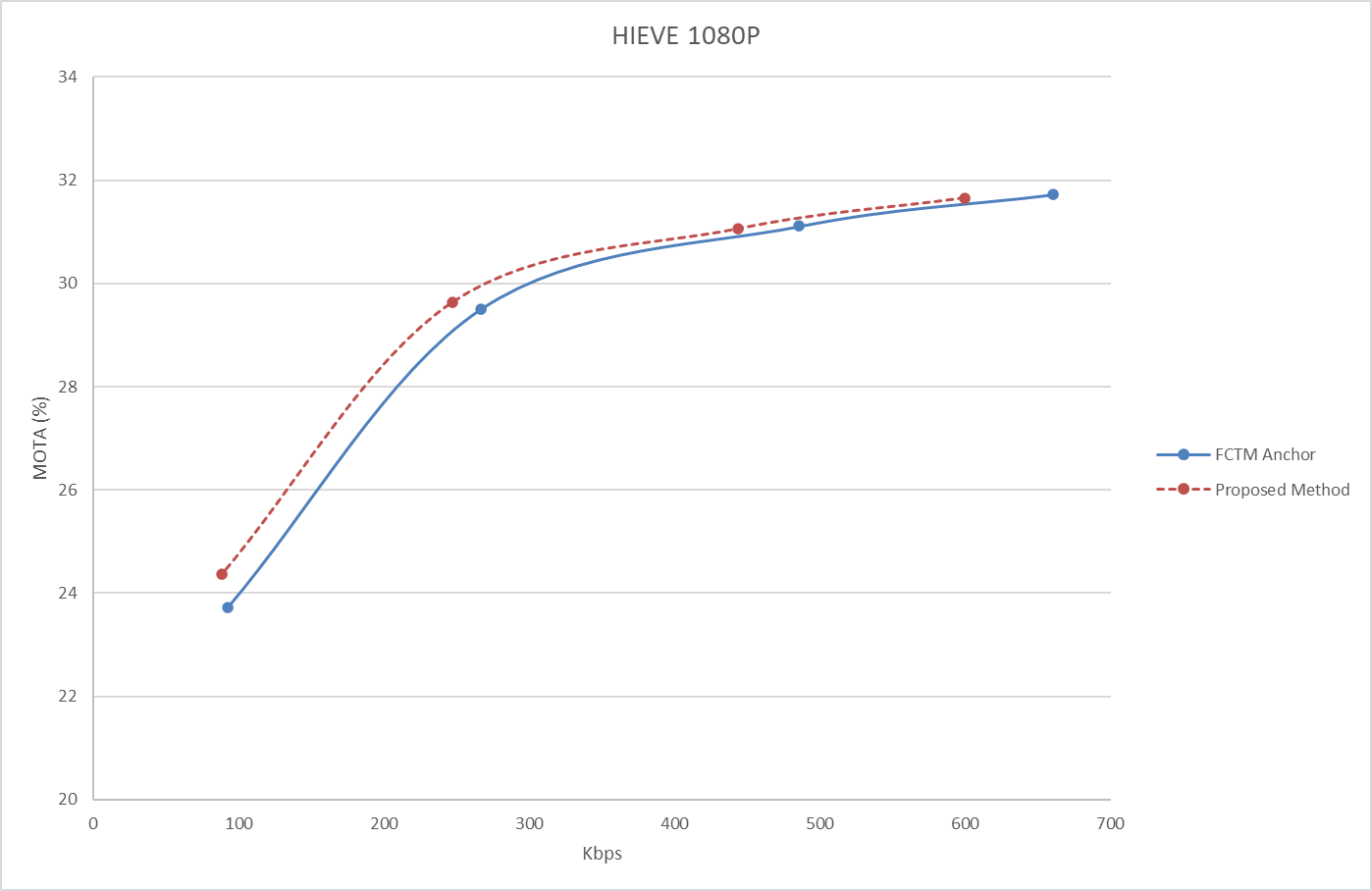}}
 \subfigure[]{\includegraphics[width=.49\linewidth]{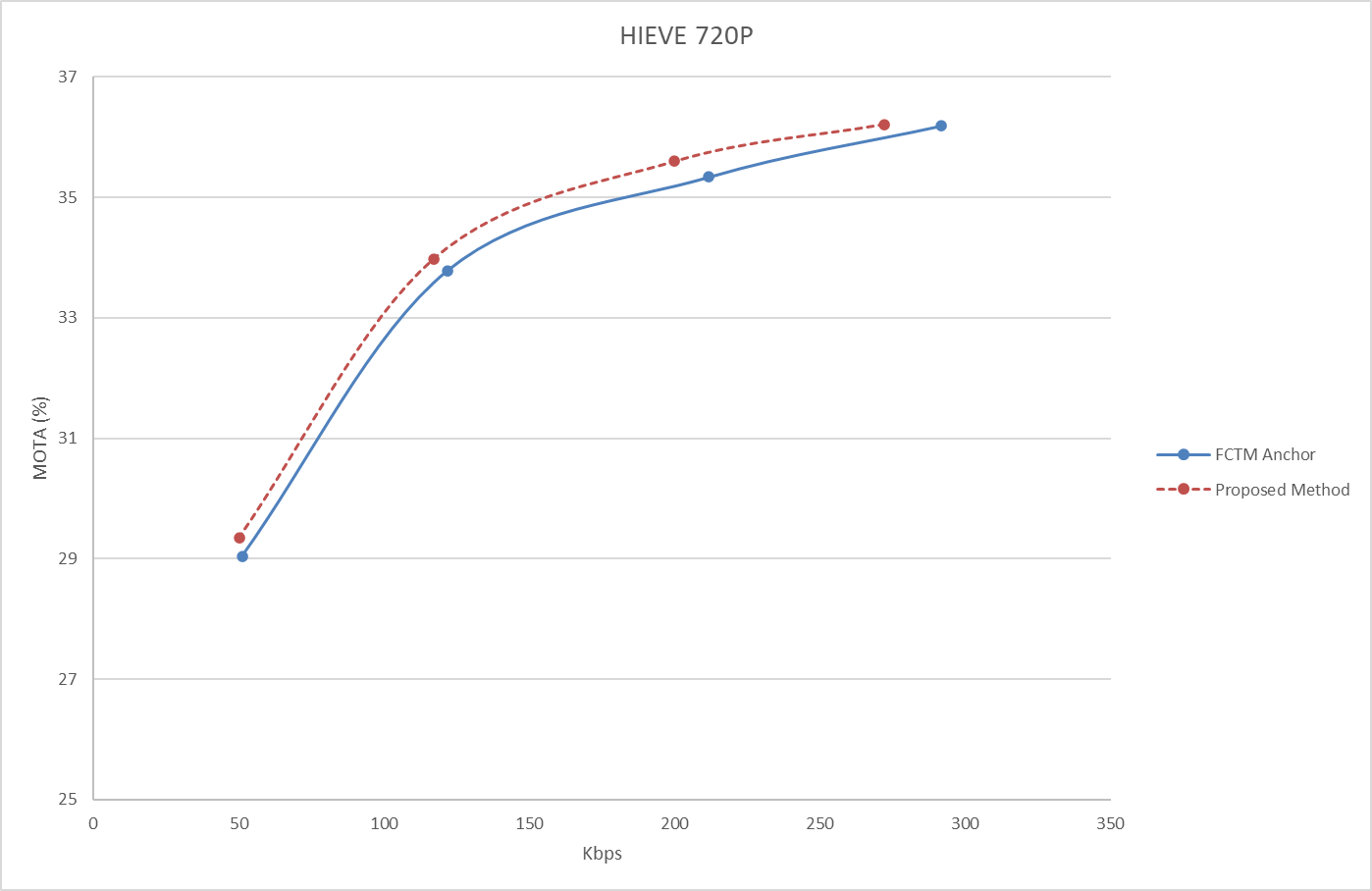}}
 \caption{HiEve bitrate vs. MOTA curves. (a) 1080p resolution, (b) 720p resolution} 
 \label{fig:hieve_curve}
\end{figure}
The results, as summarized in Table~\ref{tab:percentage_changes}, demonstrate a consistent reduction in BD-rate across multiple datasets and network split points. When compressing the intermediate features of the split detectron2 model \cite{wu2019detectron2}, tested for object detection on the SFU-HW dataset \cite{choi2021dataset}, the proposed method achieved significant BD-rate reductions. Specifically, the BD-rate decreases by 9.27\% in Class A/B while more substantial gains were observed in Class C and Class D, with reductions of 16.26\% and 18.72\%, respectively. For the TVD dataset \cite{tvd}, our method was not active due to the high number active feature channels and maintained performance.
This result demonstrates the aptitude of the proposed method to automatically deactivate channel truncation when not relevant to not penalize the overall system performance.
In the HiEve dataset, the BD-rate reduction was 11.11\% for 1080p input videos and 8.18\% for 720p input videos. Sample bitrate vs. mAP curves can be seen in Figure~\ref{fig:sfu_curve} and Figure~\ref{fig:hieve_curve}.

Overall, the proposed method led to a 10.59\% reduction in BD-rate across the evaluated datasets, indicating its effectiveness in improving video compression efficiency while introducing minimal computational overhead.
These results highlight this method's potential for improving bitrate efficiency in object detection and tracking tasks, making it a promising approach for use in real-time video applications.

\subsection{Discussion}
This method consistently achieves considerable reductions in bitrate, and the denoising effect generally enhances model accuracy compared to standalone quantization. While some dataset classes exhibit deviations in results, these are typically not directly influenced by the input resolution of the video. However, the denoising may have a more pronounced negative impact on higher-resolution video due to the increased detail captured in the feature tensors. However, this requires further study. The TVD dataset videos were not processed by our system because the feature channels were too active to remove, as detected automatically by the cutoff signal. This behavior was directly correlated with the split point, which resulted in dense activations.

\section{Conclusions} \label{sec:conc}

This paper presents a method for truncating low-activation feature channels in neural networks, specifically within the context of FCM. The proposed approach achieves significant BD-rate reductions across multiple datasets and network split points.  By dynamically identifying and removing inactive feature channels, the method optimizes the compression process with minimal computational overhead and without sacrificing task accuracy. Experimental results demonstrate the effectiveness of this method, achieving consistent BD-rate reductions across different datasets, with an overall reduction of 10.59\%. The results also indicate that the method adapts well to different network split points and input resolutions, though further investigation is needed to assess its impact on certain densely activated split points. The TVD dataset, in particular, highlighted the challenges posed by dense activations at the split point, which limited the method’s applicability.

This method introduces a promising solution for improving the efficiency of split-computing systems, especially in resource-constrained environments where reducing transmission overhead is critical. Future work, as part of a core experiment study in MPEG FCM, will focus on further refining the denoising and truncation processes, as well as utilizing information from the feature reduction and restoration methods to enable compression gains for the case of still images.

\section{References}
\bibliographystyle{IEEEbib}
\bibliography{refs}

\end{document}